\documentclass[aps,prl,showpacs,twocolumn,superscriptaddress]{revtex4}
\usepackage{bm,color}
\usepackage{graphicx}
\usepackage{amsmath}
\usepackage{color}
\pdfoutput=1

\begin{document}
\title{Resonance modes and microwave driven translational motion of skyrmion crystal under an inclined magnetic field}
\author{Masahito Ikka}
\affiliation{Department of Physics and Mathematics, Aoyama Gakuin University, Sagamihara, Kanagawa 229-8558, Japan}
\author{Akihito Takeuchi}
\affiliation{Department of Physics and Mathematics, Aoyama Gakuin University, Sagamihara, Kanagawa 229-8558, Japan}
\author{Masahito Mochizuki}
\affiliation{Department of Applied Physics, Waseda University, Okubo, Shinjuku-ku, Tokyo 169-8555, Japan}
\affiliation{PRESTO, Japan Science and Technology Agency, Kawaguchi, Saitama 332-0012, Japan}
\begin{abstract}
We theoretically investigate the microwave-active resonance modes of a skyrmion crystal on a thin-plate specimen under application of an external magnetic field that is inclined from the perpendicular direction to the skyrmion plane. In addition to the well-known breathing mode and two rotation modes, we find novel resonance modes that can be regarded as combinations of the breathing and rotation modes. Motivated by the previous theoretical work of Wang $et$ $al.$ [Phys. Rev. B {\bf 92}, 020403(R) (2015).], which demonstrated skyrmion propagation driven by breathing-mode excitation under an inclined magnetic field, we investigate the propagation of a skyrmion crystal driven by these resonance modes using micromagnetic simulations. We find that the direction and velocity of the propagation vary depending on the excited mode. In addition, it is found that a mode with a dominant counterclockwise-rotation component drives much faster propagation of the skyrmion crystal than the previously studied breathing mode. Our findings enable us to perform efficient manipulation of skyrmions in nanometer-scale devices or in magnetic materials with strong uniaxial magnetic anisotropy such as GaV$_4$S$_4$ and GaV$_4$Se$_4$, using microwave irradiation.
\end{abstract}
\maketitle
\section{Introduction}
\begin{figure}
\includegraphics[width=1.0\columnwidth]{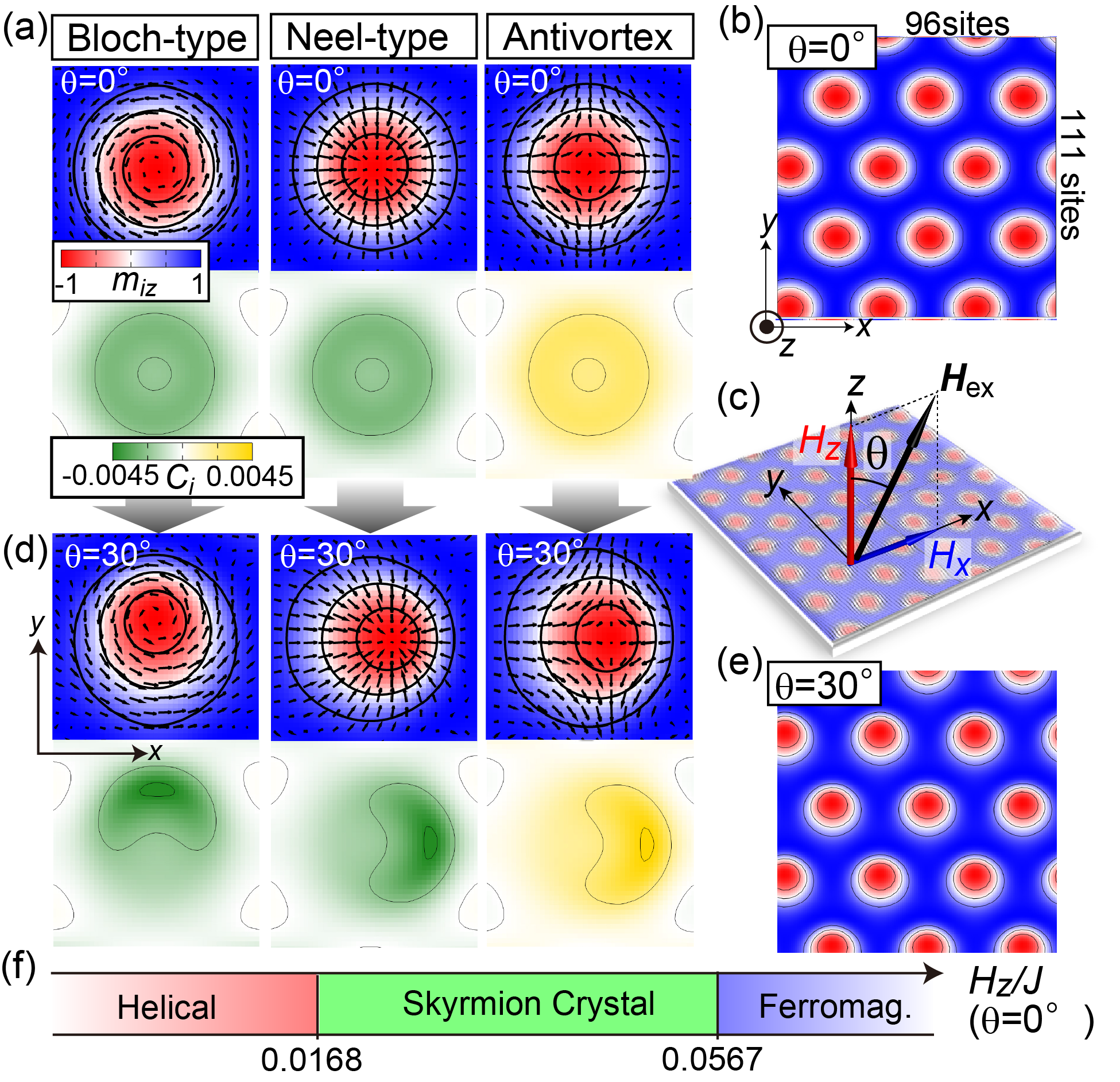}
\caption{(color online). (a) Three types of magnetic skyrmion structures. Distributions of the magnetizations (upper panels) and those of the scalar spin chiralities $C_i=(\bm m_{i+\hat{\bm x}} \times \bm m_{i+\hat{\bm x}+\hat{\bm y}})\cdot \bm m_i$ (lower panels) are shown. (b) Top view of the skyrmion crystal under a perpendicular magnetic field. (c) Thin-plate specimen hosting a skyrmion crystal under a magnetic field $\bm H_{\rm ex}$=$(H_x, 0, H_z)$ that is inclined with a finite in-plane component $H_x=H_z\tan\theta$. (d) Skyrmion structures under the inclined magnetic field. (e) Bloch-type skyrmion crystal under the inclined magnetic field. (f) Theoretical phase diagram of the spin model given by Eq.~(\ref{eqn:model}) at $T=0$ as a function of $H_z$ for a perpendicular magnetic field $\bm H_{\rm ex}$=$(0, 0, H_z)$ with $\theta=0^\circ$.}
\label{Fig01}
\end{figure}

Noncollinear spin structures in magnets such as spirals, vortices, and chiral solitons with finite helicity and/or chirality show nontrivial collective excitations and thus offer intriguing spintronics and magnonics functions. 
One of the most important examples of such spin structures is magnetic skyrmions realized in magnets with broken spatial inversion symmetry~\cite{Bogdanov89,Bogdanov94,Rossler06,Nagaosa13,Fert13,Mochizuki15a,Seki15}, in which keen competition between the Dzyaloshinskii-Moriya (DM) interaction and the ferromagnetic-exchange interaction takes place. The skyrmion structures usually appear in a plane that lies perpendicular to an external magnetic field $\bm H_{\rm ex}$, where the magnetizations at its periphery (center) are oriented parallel (antiparallel) to $\bm H_{\rm ex}$. The skyrmions are classified into three types, i.e., the Bloch-type, the Neel-type, and the antivortex-type, according to the way of their magnetization rotation [Fig.~\ref{Fig01}(a)]~\cite{Bogdanov94,Nagaosa13}.

In bulk specimens, layered skyrmion structures are stacked to form a tubular structure along the $\bm H_{\rm ex}$ direction. The skyrmions often appear in a hexagonally packed form known as a skyrmion crystal, as shown in Fig.~\ref{Fig01}(b)~\cite{Muhlbauer09,YuXZ10,Seki12,Adams12,Tonomura12}.
These skyrmions show specific microwave-active collective modes~\cite{Mochizuki12,Petrova11,LinSZ14,Schwarze15,Garst17}. When an external field $\bm H_{\rm ex}$ is applied normal to the plane of the skyrmion, the skyrmion crystal has one (two) resonance mode(s) activated by a microwave magnetic field $\bm H^\omega$ perpendicular (parallel) to the thin-plate plane. It shows a breathing mode under the perpendicular $\bm H^\omega$ field, in which the crystallized skyrmions expand and shrink uniformly in an oscillatory manner. In contrast, the two modes that occur under the in-plane $\bm H^\omega$ field are rotation modes, for which the rotational sense is counterclockwise (clockwise) for the lower-frequency (higher-frequency) mode.

Recent theoretical studies and experiments have revealed that these resonance modes of skyrmions host interesting microwave and spintronics functions~\cite{Mochizuki15a,Finocchio16}, including gigantic microwave directional dichroism~\cite{Mochizuki13,Okamura13,Mochizuki15b,Okamura15}, induction of spin voltages~\cite{Ohe13,Shimada15}, generation of spin currents~\cite{Hirobe15}, spin-torque oscillator functions~\cite{LiuRH15,ZhangS15}, microwave sensing functions~\cite{Finocchio15}, and magnonic crystal functions~\cite{MaF15,MoonKW16,Mruczkiewicz16}. These phenomena have all been investigated for the three collective modes mentioned above. Situations in which the skyrmion plane and the skyrmion tube are inclined from the $\bm H_{\rm ex}$ direction rarely occur in bulk specimens because they can easily follow the $\bm H_{\rm ex}$ direction. However, when the skyrmions are confined within a quasi-two-dimensional thin-plate specimen [see Fig.~\ref{Fig01}(c)]~\cite{Seki12,YuXZ11}, a situation can be realized in which distributions of the magnetizations and scalar spin chiralities have disproportionate weight and are slanted from the skyrmion center, as shown in Fig.~\ref{Fig01}(d) and (e)~\cite{LinSZ15}. Such a situation can also occur in magnets with strong uniaxial magnetic anisotropies, in which the orientations of the skyrmion plane and the skyrmion tubes are fixed, irrespective of the $\bm H_{\rm ex}$ direction.

We expect emergence of characteristic resonance modes for skyrmion crystal under application of an inclined $\bm H_{\rm ex}$ field. However, such modes have not been studied systematically to date, although novel modes are expected to host previously unrecognized functions and phenomena. Indeed, a recent theoretical study proposed that translational motion of skyrmions can be driven by application of a microwave magnetic field $\bm H^\omega$ to a skyrmion-hosting two-dimensional system under an inclined $\bm H_{\rm ex}$ field~\cite{WangW15}. In addition, it was found that a skyrmion crystal appears always on the (111)-plane, irrespective of the $\bm H_{\rm ex}$ direction in insulating vanadates GaV$_4$S$_8$ and GaV$_4$Se$_8$ with lacunar spinel structure because of their strong uniaxial magnetic anisotropy~\cite{Kezsmarki15,Ehlers16,Ehlers17}. In insulating skyrmion-hosting materials of this type, specific resonance modes can be sources of interesting microwave magnetoelectric phenomena owing to their multiferroic nature with magnetically induced electric polarizations. Under these circumstances, clarification of the microwave-active modes and the microwave-related phenomena of skyrmions under application of an inclined $\bm H_{\rm ex}$ field becomes an issue of major importance.

In this paper, we theoretically investigate the microwave-active resonance modes of a skyrmion crystal in a two-dimensional system under an inclined $\bm H_{\rm ex}$ field. By numerically solving the Landau-Lifshitz-Gilbert (LLG) equation, we trace dynamics of the magnetizations that constitute the skyrmion crystal to calculate microwave absorption spectra and obtain real-space snapshots for each eigenmode. In addition to the well-known breathing and two types of rotation modes, we find that characteristic modes appear, which can be regarded as combinations of the breathing and rotation modes. Using micromagnetic simulations, we demonstrate that continuous excitation of these resonance modes via microwave application results in propagation of the skyrmion crystal where its direction and velocity sensitively depend on the excited mode (or the microwave frequency) and the microwave intensity. Furthermore, we find that a mode with a dominant counterclockwise-rotation component drives much faster propagation of the skyrmion crystal than the previously examined breathing mode. The knowledge of these resonance modes and the microwave-driven motion of skyrmion crystals under an inclined magnetic field lead to techniques to manipulate skyrmions using microwaves and provide a means to realize unique skyrmion-based devices.

\section{Spin Model and Method}
We employ a classical Heisenberg model on a square lattice to describe the magnetism in a thin-plate specimen of a skyrmion-hosting magnet, which contains the ferromagnetic exchange interaction, the DM interaction among the normalized magnetization vectors $\bm m_i$, and the Zeeman coupling to the external magnetic field $\bm H_{\rm ex}$~\cite{Bak80,YiSD09}. The Hamiltonian is given by,
\begin{eqnarray}
\mathcal{H}_0&=&
-J \sum_{<i,j>} \bm m_i \cdot \bm m_j
+\sum_{i,\hat{\bm \gamma}} \bm D_\gamma \cdot (
\bm m_i \times \bm m_{i+\hat{\bm \gamma}})
\nonumber \\
&-&\bm H_{\rm ex} \cdot \sum_i \bm m_i,
\label{eqn:model}
\end{eqnarray}
Types of the skyrmion are determined by a structure of the Moriya vectors $\bm D_\gamma$ ($\gamma=x, y$), i.e., $\bm D_x=(D, 0)$ and $\bm D_y=(0, D)$ produce the Bloch-type skyrmion, $\bm D_x=(0, D)$ and $\bm D_y=(-D, 0)$ produce the Neel-type skyrmion, and $\bm D_x=(0, D)$ and $\bm D_y=(D, 0)$ produce the antivortex-type skyrmion. We adopt $J$=1 for the energy units and set $D/J$=0.27. The external magnetic field $\bm H_{\rm ex}$ is inclined from the perpendicular direction ($\parallel$$\bm z$) towards the $x$ direction as $\bm H_{\rm ex}$=$(H_x, 0, H_z)$ with $H_x=H_z\tan\theta$, where $\theta$ is the inclination angle [see Fig.~\ref{Fig01}(c)]. Figure~\ref{Fig01}(f) shows a theoretical phase diagram of this spin model at $T$=0 as a function of $H_z$ when $\bm H_{\rm ex}$ is applied normal to the two-dimensional plane ($\theta$=0). This phase diagram exhibits the skyrmion-crystal phase in a region of moderate field strength sandwiched by the helical phase and the field-polarized ferromagnetic phase. The unit conversions when $J$=1 meV are summarized in Table~\ref{tab:uconv}.
\begin{table}
\begin{tabular}{l|cc} \hline \hline
Exchange int.  & \hspace{0.5cm} $J=1$             & \hspace{0.5cm} 1 meV \\
Time           & \hspace{0.5cm} $t=1$             & \hspace{0.5cm} 0.66 ps \\
Frequency $f=\omega/2\pi$ &\hspace{0.5cm} $\omega=1$ & \hspace{0.5cm} 241 GHz\\
               & \hspace{0.5cm} ($\omega=0.01$ & \hspace{0.5cm} 2.41 GHz) \\
Magnetic field & \hspace{0.5cm} $H=1$             & \hspace{0.5cm} 8.64 T \\ 
\hline \hline
\end{tabular}
\caption{Unit conversion table when $J$=1 meV.}
\label{tab:uconv}
\end{table}

\begin{figure*}
\includegraphics[width=2.0\columnwidth]{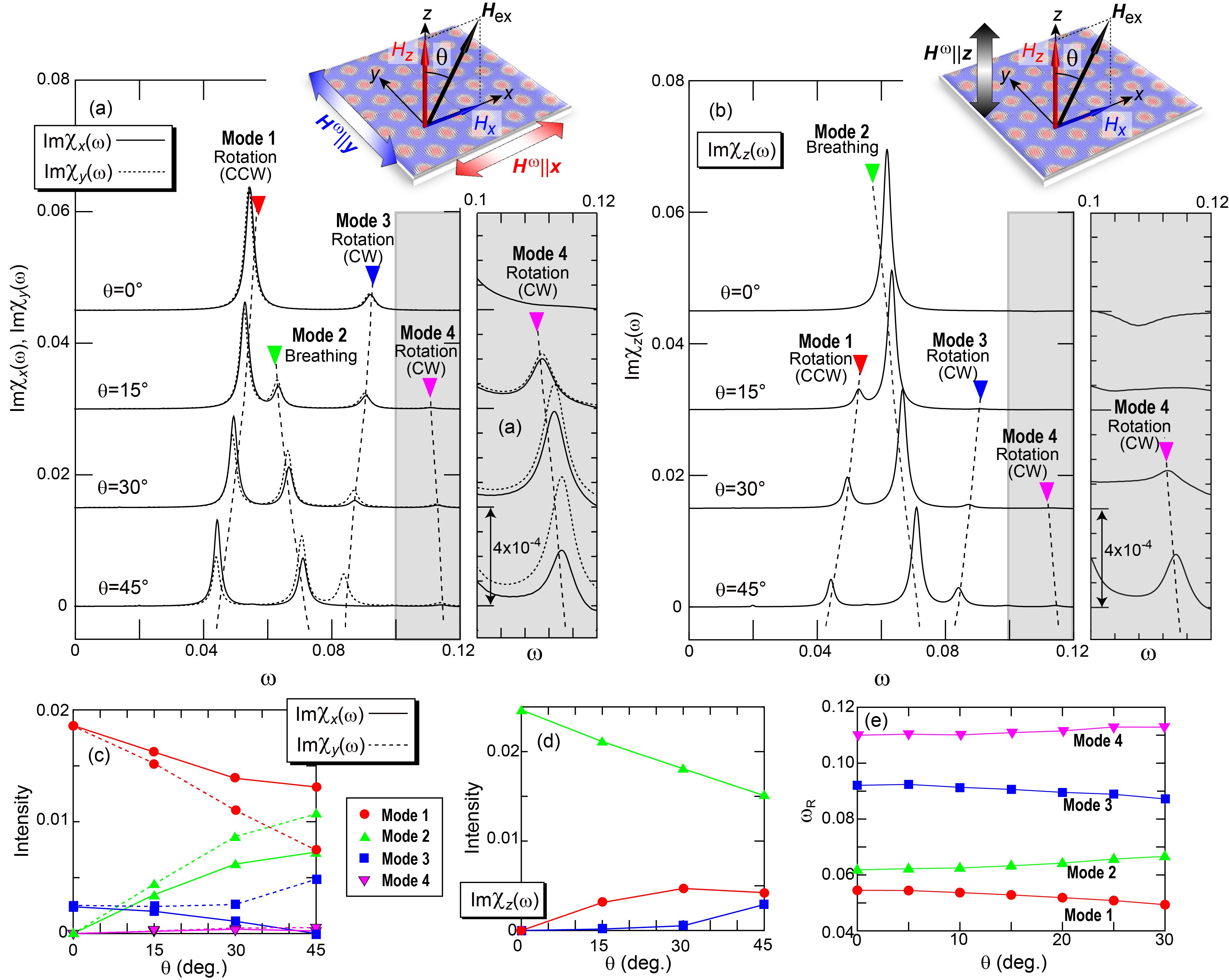}
\caption{(color online). (a), (b) Imaginary parts of the calculated dynamical magnetic susceptibilities of skyrmion crystal confined in a two-dimensional system under perpendicular ($\theta$=0$^{\circ}$) and inclined ($\theta\ne$0$^{\circ}$) magnetic fields as functions of the angular frequency $\omega$. (a) Those for the in-plane microwave polarization with $\bm H^\omega$$\parallel$$\bm x$,$\bm y$. (b) Those for the out-of-plane microwave polarization with $\bm H^\omega$$\parallel$$\bm z$. Note that these spectra are calculated for the Bloch-type skyrmion crystal, but it was confirmed that the Neel-type and the antivortex-type skyrmion crystals have perfectly equivalent spectra. Here the inclined magnetic field is given by $\bm H_{\rm ex}$=$(H_z\tan\theta, 0, H_z)$ with $H_z$=0.036. A dominant component of the oscillation is indicated below the name of each mode, where CW and CCW indicate the clockwise and the counterclockwise rotations, respectively. The spectra for Mode 4 are magnified in the insets. (c), (d) Spectral intensities as functions of $\theta$ for Modes 1-4 for (c) $\bm H^\omega$$\parallel$$\bm x$,$\bm y$ and (d)$\bm H^\omega$$\parallel$$\bm z$. (e) Resonant frequencies $\omega_{\rm R}$ as functions of $\theta$ for Modes 1-4.}
\label{Fig02}
\end{figure*}
\begin{figure*}
\includegraphics[width=2.0\columnwidth]{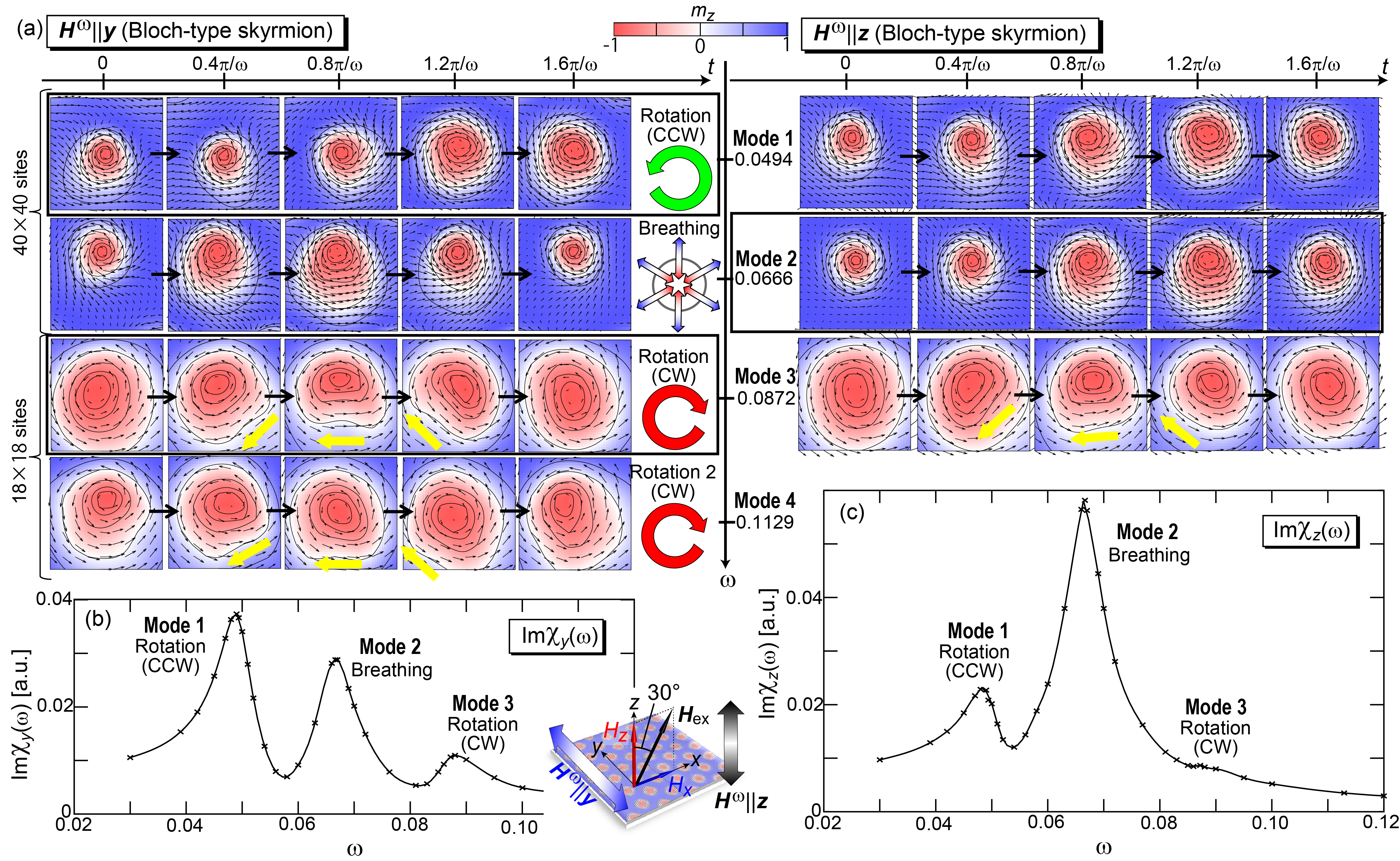}
\caption{(color online). (a) Snapshots of the four resonance modes (Modes 1-4) of skyrmion crystal activated by the in-plane microwave field $\bm H^\omega$$\parallel$$\bm y$ (left panels) and those of the three resonance modes (Modes 1-3) activated by the out-of-plane microwave field $\bm H^\omega$$\parallel$$\bm z$ (right panels) under an inclined magnetic field $\bm H_{\rm ex}$=$(H_z\tan\theta, 0, H_z)$ with $H_z$=0.036 and $\theta$=30$^{\circ}$. One skyrmion in the skyrmion crystal is focused on because all the skyrmions in the skyrmion crystal show identical motions. We show here the results for the Bloch-type skyrmion crystal, but the other two skyrmion types show equivalent behaviors, with the exception of the rotation sense of the antivortex-type (see text and Fig.~\ref{Fig04}). (b), (c) Imaginary parts of the dynamical magnetic susceptibilities calculated by tracing the time-profiles of net magnetization through applying a sinusoidal AC magnetic field $\bm H(t)$ for (b) $\bm H^\omega$$\parallel$$\bm y$ and (c) $\bm H^\omega$$\parallel$$\bm z$. Both spectra exhibit peaks at frequencies equivalent to those of the spectra for $\theta=30^\circ$ shown in Fig.~\ref{Fig02}, which were calculated by applying a short rectangular pulse. This indicates that both the short pulse and the AC field excite identical modes and validates our method to study the modes.}
\label{Fig03}
\end{figure*}
\begin{figure}
\includegraphics[width=1.0\columnwidth]{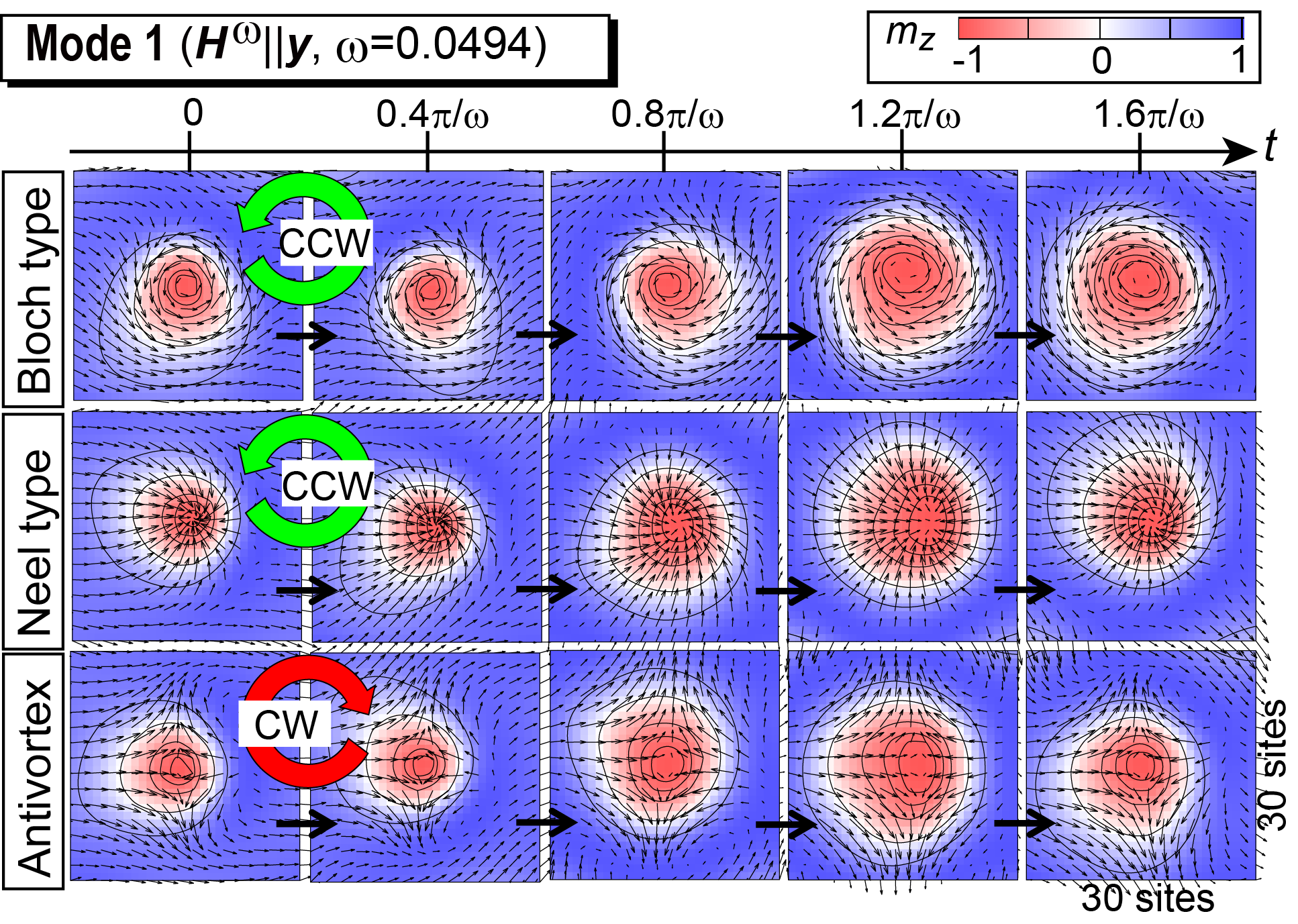}
\caption{(color online). Snapshots of Mode 1 for Bloch-type, Neel-type and antivortex-type skyrmion crystals under an inclined magnetic field $\bm H_{\rm ex}$=$(H_z\tan\theta, 0, H_z)$ with $H_z$=0.036 and $\theta$=30$^{\circ}$ activated by the in-plane microwave field $\bm H^\omega$$\parallel$$\bm y$. Note that the rotation sense of the antivortex type is opposite to those of the Bloch type and the Neel type.}
\label{Fig04}
\end{figure}
The LLG equation is given by,
\begin{equation}
\frac{d\bm m_i}{dt}=-\gamma \bm m_i \times \bm H^{\rm eff}_i 
+\frac{\alpha_{\rm G}}{m} \bm m_i \times \frac{d\bm m_i}{dt}.
\label{eq:LLGEQ}
\end{equation} 
Here $\alpha_{\rm G}$(=0.02) is the Gilbert-damping coefficient. The effective magnetic field $\bm H_i^{\rm eff}$ acting on the local magnetization $\bm m_i$ on the $i$-th site is calculated from the Hamiltonian $\mathcal{H}$=$\mathcal{H}_0$+$\mathcal{H}'(t)$ in the form
\begin{equation}
\bm H^{\rm eff}_i = -\frac{1}{\gamma\hbar}\frac{\partial \mathcal{H}}{\partial \bm m_i}.
\label{eq:EFFMF}
\end{equation}
The first term, $\mathcal{H}_0$, is the model Hamiltonian given by Eq.~(\ref{eqn:model}). The second term, $\mathcal{H}'(t)$, represents the coupling between the magnetizations and a time-dependent magnetic field $\bm H(t)$ in the form
\begin{equation}
\mathcal{H}'(t)=-\bm H(t) \cdot \sum_i \bm m_i.
\label{eq:tdepH}
\end{equation}
The calculations are performed using a system of $N=96 \times 111$ sites where the periodic boundary condition is imposed.

\section{Results for the Resonance Modes}
To identify the resonance modes of a skyrmion crystal under application of an inclined $\bm H_{\rm ex}$ field, we first calculate dynamical magnetic susceptibilities,
\begin{eqnarray}
\chi_\alpha(\omega) =
\frac{\Delta M^\alpha(\omega)}{H^\alpha(\omega)} \quad\quad
(\alpha=x, y, z),
\end{eqnarray}
Here $H^\alpha(\omega)$ and $\Delta M^\alpha(\omega)$ are Fourier transforms of the time-dependent magnetic field $\bm H(t)$ and the simulated time-profile of the total magnetization $\Delta \bm M(t)=\bm M(t)-\bm M(0)$ with $\bm M(t)=\frac{1}{N}\sum_{i=1}^N \bm m_i(t)$. For these calculations, we use a short rectangular pulse for the time-dependent field $\bm H(t)$ whose components are given by,
\begin{eqnarray}
H^\alpha(t)=\left\{
\begin{aligned}
& H_{\rm pulse} \quad & 0 \le t \le 1 \\
& 0 \quad & {\rm others}
\end{aligned}
\right.
\end{eqnarray}
where $t=(J/\hbar)\tau$ is the dimensionless time with $\tau$ being the real time. An advantage of using the short pulse is that for a sufficiently short duration $\Delta t$ with $\omega \Delta t \ll 1$, the Fourier component $H^\alpha(\omega)$ becomes constant being independent of $\omega$ up to the first order of $\omega \Delta t$. The Fourier component is calculated as 
\begin{eqnarray}
H^\alpha(\omega)&=&\int_0^{\Delta t}\;H_{\rm pulse} e^{i\omega t}dt
=\frac{H_{\rm pulse}}{i\omega}\left(e^{i\omega \Delta t}-1 \right)
\nonumber \\
&\sim&H_{\rm pulse} \Delta t.
\end{eqnarray}
As a result, we obtain the relationship $\chi_\alpha(\omega) \propto \Delta M^\alpha(\omega)$.

In Fig~\ref{Fig02}(a), we present calculated microwave absorption spectra, i.e., imaginary parts of the dynamical magnetic susceptibilities Im$\chi_x$ and Im$\chi_y$ for the in-plane microwave fields $\bm H^\omega$$\parallel$$\bm x$ and $\bm H^\omega$$\parallel$$\bm y$, respectively, as functions of microwave frequency $\omega(=2\pi f)$ for several values of inclination angle $\theta$. Here, we fix $H_z$=0.036. Note that while these spectra are calculated for the Bloch-type skyrmion crystal, we confirmed that the Neel-type and the antivortex-type show perfectly equivalent spectra. We find that while only two rotation modes with counterclockwise and clockwise rotation senses (referred to as Modes 1 and 3, respectively) are active under the perpendicular $\bm H_{\rm ex}$ field with $\theta$=0$^{\circ}$, novel modes (Modes 2 and 4) appear when the $\bm H_{\rm ex}$ field is inclined with $\theta\ne0$. Mode 2 appears between the two original rotation modes (Modes 1 and 3) in frequency, whereas Mode 4 has a higher frequency than Mode 3. 

Intensities of these novel modes grow as the inclination angle $\theta$ increases. As will be discussed below, Modes 2 and 4 can be regarded as combinations of the breathing and clockwise rotation modes, where the former (latter) component is dominant for Mode 2 (Mode 4). On the other hand, intensity of Mode 3 is either enhanced or suppressed with increasing $\theta$, depending on the polarization of the microwave field $\bm H^\omega$. When the microwave field $\bm H^\omega$ is oriented parallel (perpendicular) to the direction toward which the $\bm H_{\rm ex}$ field is inclined, i.e., $\bm H^\omega$$\parallel$$\bm x$ ($\bm H^\omega$$\parallel$$\bm y$), the intensity becomes suppressed (enhanced) as $\theta$ increases.

Figure~\ref{Fig02}(b) shows imaginary parts of the dynamical magnetic susceptibilities Im$\chi_z$ calculated for the out-of-plane microwave field $\bm H^\omega$$\parallel$$\bm z$ for several values of $\theta$. Again, we find that novel modes (Modes 1, 3 and 4) appear when the $\bm H_{\rm ex}$ field is inclined, whereas the single breathing mode (Mode 2) alone exists under the perpendicular $\bm H_{\rm ex}$ field. Intensities of these novel modes increase as $\theta$ increases, while intensity of the original Mode 2 decreases. As will be discussed below, Mode 1 (Modes 3 and 4) can be regarded as combined oscillations of the dominant counterclockwise (clockwise) rotation component and the subsequent breathing component. 

Comparison of the calculated dynamical magnetic susceptibilities in Fig.~\ref{Fig02}(a) and (b) shows that four types of collective modes activated by $\bm H^\omega$$\parallel$$\bm x$, $y$ have identical resonance frequencies with four corresponding collective modes activated by $\bm H^\omega$$\parallel$$\bm z$, indicating that both the in-plane microwave fields $\bm H^\omega$$\parallel$$\bm x$, $\bm y$ and the out-of-plane microwave field $\bm H^\omega$$\parallel$$\bm z$ activate equivalent modes under application of the inclined $\bm H_{\rm ex}$ field. 

In Fig.~\ref{Fig03}, snapshots of the magnetization distributions for each mode focusing on a skyrmion constituting the Bloch-type skyrmion crystal are shown for $\bm H^\omega$$\parallel$$\bm y$ (left panels) and $\bm H^\omega$$\parallel$$\bm z$ (right panels) when $H_z$=0.036 and $\theta$=30$^\circ$. They are simulated via application of a microwave magnetic field $H^\alpha(t)=H_\alpha^\omega \sin\omega t$ ($\alpha$=$x,y,z$) with a corresponding resonance frequency as the time-dependent magnetic field $\bm H(t)$ in Eq.~(\ref{eq:tdepH}). In the simulations of these eigenmode dynamics, we have monitored time profiles of net magnetization and their Fourier transforms to confirm that a pure eigenmode with a single-frequency component is excited while other modes are absent. Noticeably, Mode 2 under $\bm H^\omega$$\parallel$$\bm y$ and Mode 2 under $\bm H^\omega$$\parallel$$\bm z$ at $\omega$=0.0666 are identical, which are regarded as a breathing mode. It is also found that Mode 1 (Mode 3) under $\bm H^\omega$$\parallel$$\bm y$ and Mode 1 (Mode 3) under $\bm H^\omega$$\parallel$$\bm z$ at $\omega$=0.0494 ($\omega$=0.0872) are again identical, and can be regarded as combined oscillations of the dominant counterclockwise (clockwise) rotation and the subsequent breathing component. While the absorption intensity is too weak to enable snapshots of Mode 4 to be obtained under $\bm H^\omega$$\parallel$$\bm z$, we expect that they would be equivalent to those of Mode 4 under $\bm H^\omega$$\parallel$$\bm y$. 

It is known that a single skyrmion in a constricted geometry exhibits quantized higher harmonic radial and azimuthal spin-wave modes when the system is activated by a sinusoidal AC field, especially in the presence of relatively strong damping effects~\cite{Kim14,Beg17}. In the present calculations, we used an infinite and uniform system with the periodic boundary conditions, and thus such higher harmonics should be absent. However, it might be important to check their absence. For this purpose, we calculate the dynamical magnetic susceptibilities by applying a sinusoidal AC magnetic field. Figure~\ref{Fig03}(b) [(c)] shows the calculated spectrum for $\bm H^\omega$$\parallel$$\bm y$ [$\bm H^\omega$$\parallel$$\bm z$] for $\theta=30^\circ$, which has peaks at the same frequencies with that in Fig.~\ref{Fig02}, indicating the absence of higher harmonic modes and the validity of the identified eigenmodes. Note that calculations of the spatial distributions of power and phase may provide useful information to identify the eigenmodes~\cite{Kumar11}.

In addition to the resonance modes of the Bloch-type skyrmion crystal, we also examined those of the Neel-type and the antivortex-type skyrmion crystals. We find that the microwave absorption spectra for these three different types of skyrmion crystals overlap perfectly, indicating that the three skyrmion crystals have resonance modes with identical frequencies and identical intensities. However, differences appear in terms of the rotation senses of Modes 1, 3, and 4. The rotation sense of the antivortex skyrmion crystal is always opposite to that of the corresponding modes of the Bloch-type and Neel-type skyrmion crystals. Figure~\ref{Fig04} shows snapshots of the calculated magnetization distributions of Mode 1 for the three types of skyrmion crystals. The rotation sense is counterclockwise for the Bloch-type and the Neel-type, whereas it is clockwise for the antivortex-type~\cite{Nayak17}.

To close this section, it is worth mentioning that the present study is based on a pure two-dimensional model although real specimens have finite thickness. In a real specimen, magnetic structures at the surfaces might be different from those inside the specimen because magnetizations at the surfaces have neighbors only on one side~\cite{Rybakov16,ZhangSL18}. Such surface magnetic structures might affect the properties of magnetic resonances quantitatively. However, we expect that our results will not be changed qualitatively or semi-quantitatively even if we adopt a three-dimensional model. In addition, the demagnetization effects due to the magnetic dipole interactions are not incorporated in the present study. A recent study on the microwave-active skyrmion resonances revealed that frequencies and amplitudes of the skyrmion modes vary depending on the shape and thickness of the specimens due to the demagnetization effects~\cite{Schwarze15}. However, the results of Ref.~\cite{Schwarze15} also indicated that the demagnetization effects do not change the properties of resonant modes qualitatively.

\section{Results of Translational Motion}
\begin{figure}
\includegraphics[width=1.0\columnwidth]{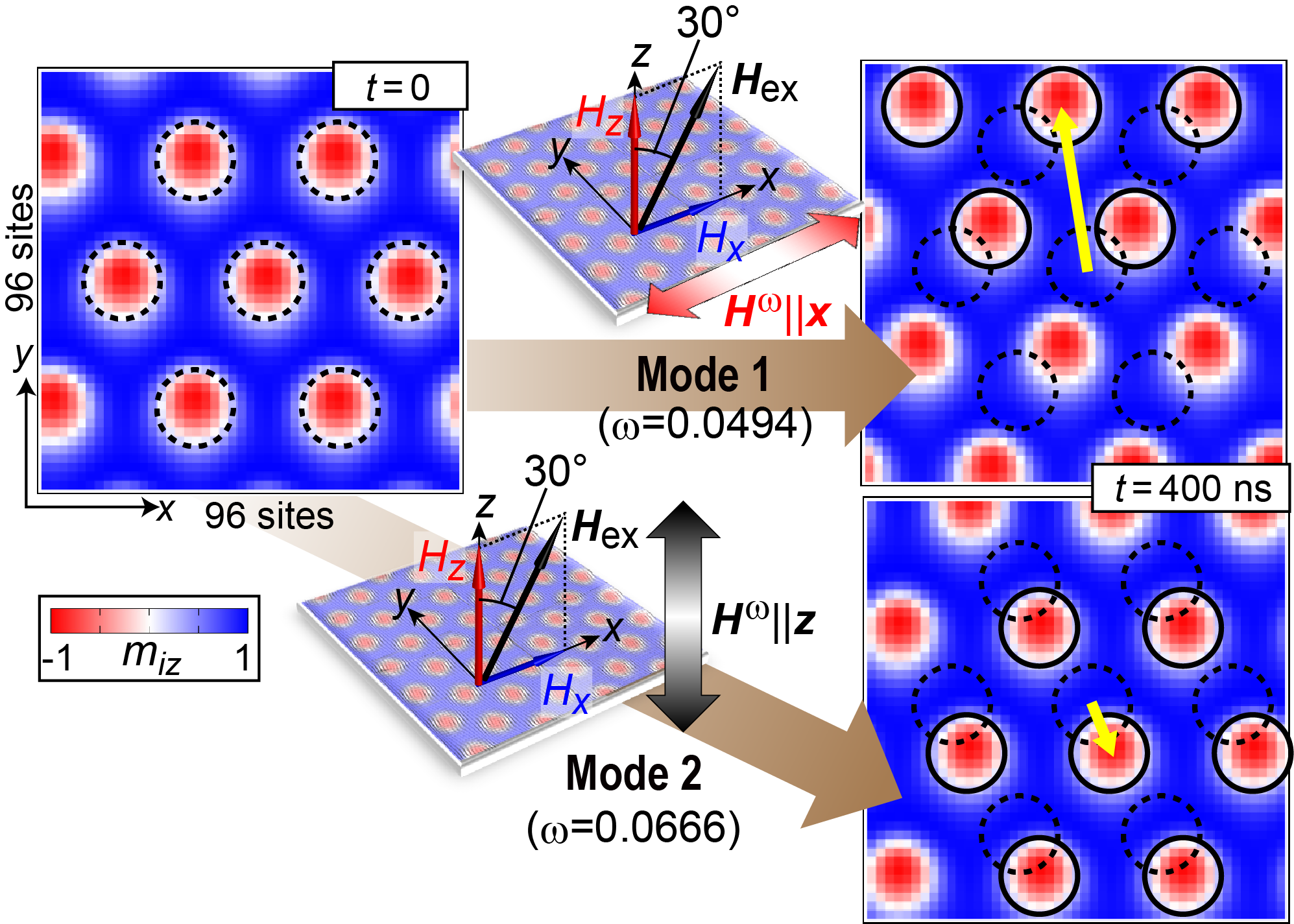}
\caption{(color online). Translational motion of Bloch-type skyrmion crystal in a thin-plate specimen driven by microwave irradiation through activation of the resonance modes under an inclined magnetic field $\bm H_{\rm ex}$=$(H_z\tan\theta, 0, H_z)$ with $H_z$=0.036 and $\theta$=30$^{\circ}$. Here, the microwave field is given by $H^\omega_\alpha \sin\omega t$ ($\alpha=x,y$) with $H^\omega_\alpha=0.0006$. The skyrmion crystal moves approximately towards the positive (negative) $y$ direction when a microwave field $\bm H^\omega$$\parallel$$\bm x$ ($\bm H^\omega$$\parallel$$\bm z$) with $\omega=0.0494$ ($\omega=0.0666$) activates Mode 1 (Mode 2) with a dominant counterclockwise-rotation (breathing) component. Displacement vectors connecting the original position (dashed circles) and the position after microwave irradiation for 400 ns (solid circles) are indicated by the thick arrows in the right panels. The motion driven by Mode 1 is much faster than that driven by Mode 2.}
\label{Fig05}
\end{figure}
A recent theoretical study by Wang $et$ $al.$ discovered that translational motion of skyrmions can be driven by an out-of-plane microwave field $\bm H^\omega$$\parallel$$\bm z$ under an inclined magnetic field $\bm H_{\rm ex}$ through activation of their breathing oscillations~\cite{WangW15}. Motivated by this study, we investigate the motions of a two-dimensional skyrmion crystal driven by several different resonance modes under the $\bm H_{\rm ex}$ field inclined towards the $x$ direction. In the numerical simulations, we find that the translational motion can be driven not only by the previously examined out-of-plane microwave field $\bm H^\omega$$\parallel$$\bm z$ but also by in-plane microwave fields $\bm H^\omega$$\parallel$$x$, $y$ via activation of the rotational oscillations of the skyrmions. 

Figure~\ref{Fig05} shows snapshots of the skyrmion crystal driven by $\bm H^\omega$$\parallel$$\bm x$ (right upper panel) and the same skyrmion crystal driven by $\bm H^\omega$$\parallel$$\bm z$ (right lower panel) at $t$=400 ns after the microwave irradiation commences. The figure also shows the initial configuration of the skyrmion crystal at $t$=0 (left panel) under application of an inclined magnetic field $\bm H_{\rm ex}$=$(H_z\tan\theta, 0, H_z)$ with $H_z$=0.036 and $\theta$=30$^{\circ}$. Here, the microwave field is given by $H^\omega_\alpha \sin\omega t$ ($\alpha=x,y$) with $H^\omega_\alpha=0.0006$. An area composed of 96$\times$96 is magnified in the figure, although the simulations are performed using a system of 96$\times$111 sites. The displacement vectors connecting the original position and the position at $t$=400 ns are indicated by the thick arrows shown in the right-side panels. When the microwave field $\bm H^\omega$$\parallel$$\bm x$ with $\omega$=0.0494 activates Mode 1 with a dominant counterclockwise-rotation component, the skyrmion crystal propagates in a direction close to the positive $y$ direction, whereas the same skyrmion crystal propagates in a direction close to the negative $y$ direction when $\bm H^\omega$$\parallel$$\bm z$ with $\omega=0.0666$ activates Mode 2 with breathing oscillations. We also find that the travel distance in the former case is much longer than that in the latter case, which indicates that the in-plane microwave field $\bm H^\omega$$\parallel$$x$ drives much faster motions of the skyrmion crystal than the out-of-plane microwave field $\bm H^\omega$$\parallel$$\bm z$.

\begin{figure}
\includegraphics[width=1.0\columnwidth]{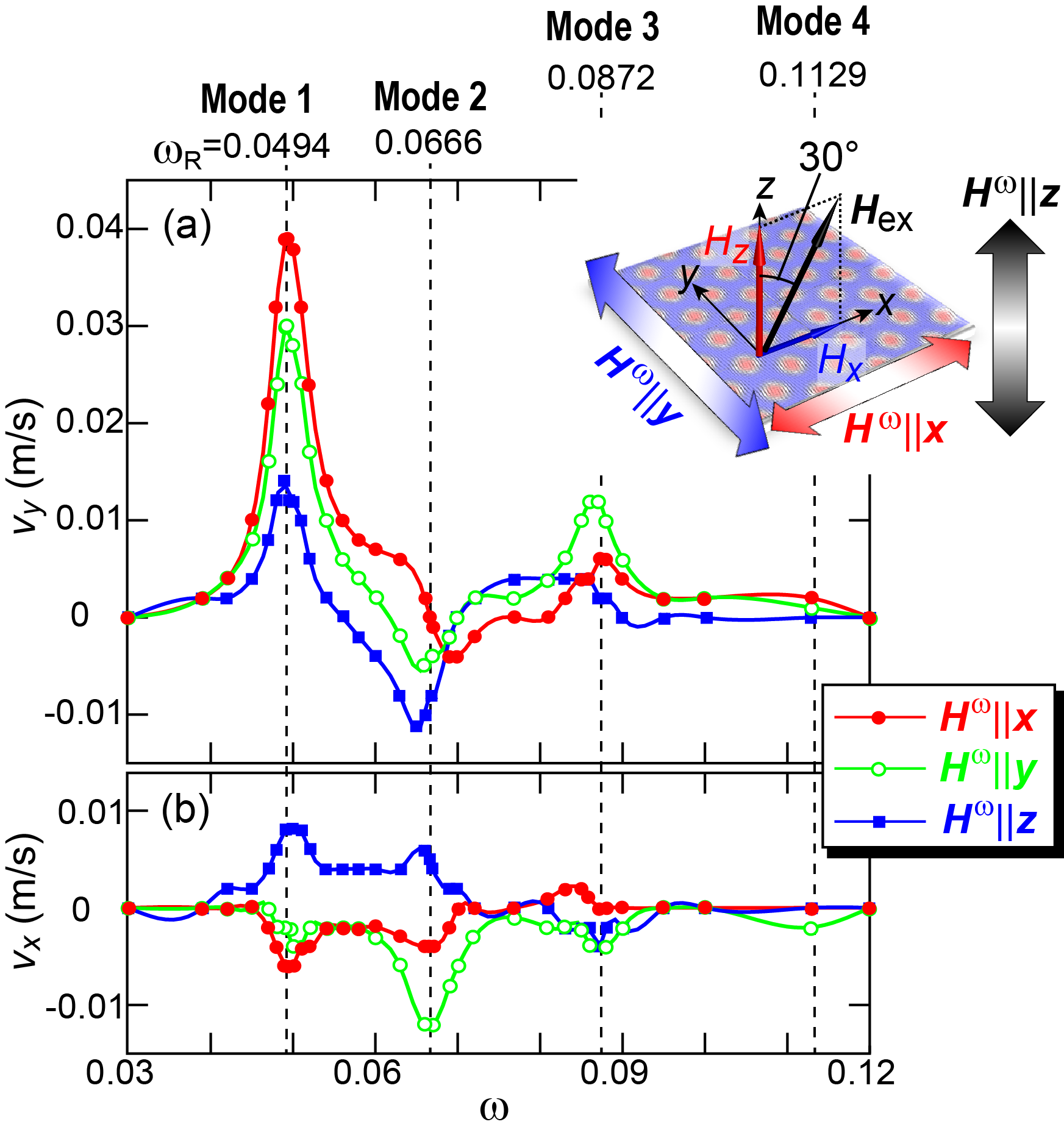}
\caption{(color online). Calculated $\omega$-dependence of the velocity $\bm v=(v_x,v_y)$ for translational motion of the Bloch-type skyrmion crystal induced by a microwave field $\bm H^\omega$ under an inclined magnetic field $\bm H_{\rm ex}$=$(H_z\tan\theta, 0, H_z)$ with $H_z$=0.036 and $\theta$=30$^{\circ}$. Here the microwave field is given by $H^\omega_\alpha \sin\omega t$ ($\alpha=x,y$) with $H^\omega_\alpha$=0.0006, and $\omega=2\pi f$ is its angular frequency. The velocities show peaks at the resonant frequencies of the modes, while their signs vary depending on the mode.}
\label{Fig06}
\end{figure}
Next we investigate the microwave frequency dependence of the velocity $\bm v$=$(v_x,v_y)$ of the driven skyrmion crystal under an inclined magnetic field $\bm H_{\rm ex}$=$(H_z\tan\theta, 0, H_z)$ with $H_z$=0.036 and $\theta$=30$^{\circ}$. In Figs.~\ref{Fig06}(a) and (b), we plot the simulated $\omega$-dependence of $v_x$ and $v_y$, respectively, for different microwave polarizations where the microwave amplitude is set to be $H_\alpha^\omega$=0.0006. We find that the velocities are enhanced to have peaks at frequencies that correspond to the resonant modes, whereas their signs vary depending on the mode. We also find that the velocity is highest when the in-plane microwave field $\bm H^\omega$$\parallel$$\bm x$ ($\omega=0.0494$) activates Mode 1 with the dominant counterclockwise-rotation component. In this case, the value of $v_x$ becomes $\sim$0.04 m/s. In contrast, the velocity when Mode 2 is activated at $\omega=0.666$ is highest in the case of the out-of-plane microwave field $\bm H^\omega$$\parallel$$\bm z$, where it reaches $-0.01$ m/s. Note that the in-plane microwave field $\bm H^\omega$$\parallel$$\bm x$ drives the skyrmion crystal approximately four times faster than the out-of-plane microwave field $\bm H^\omega$$\parallel$$\bm z$.

\begin{figure}
\includegraphics[width=1.0\columnwidth]{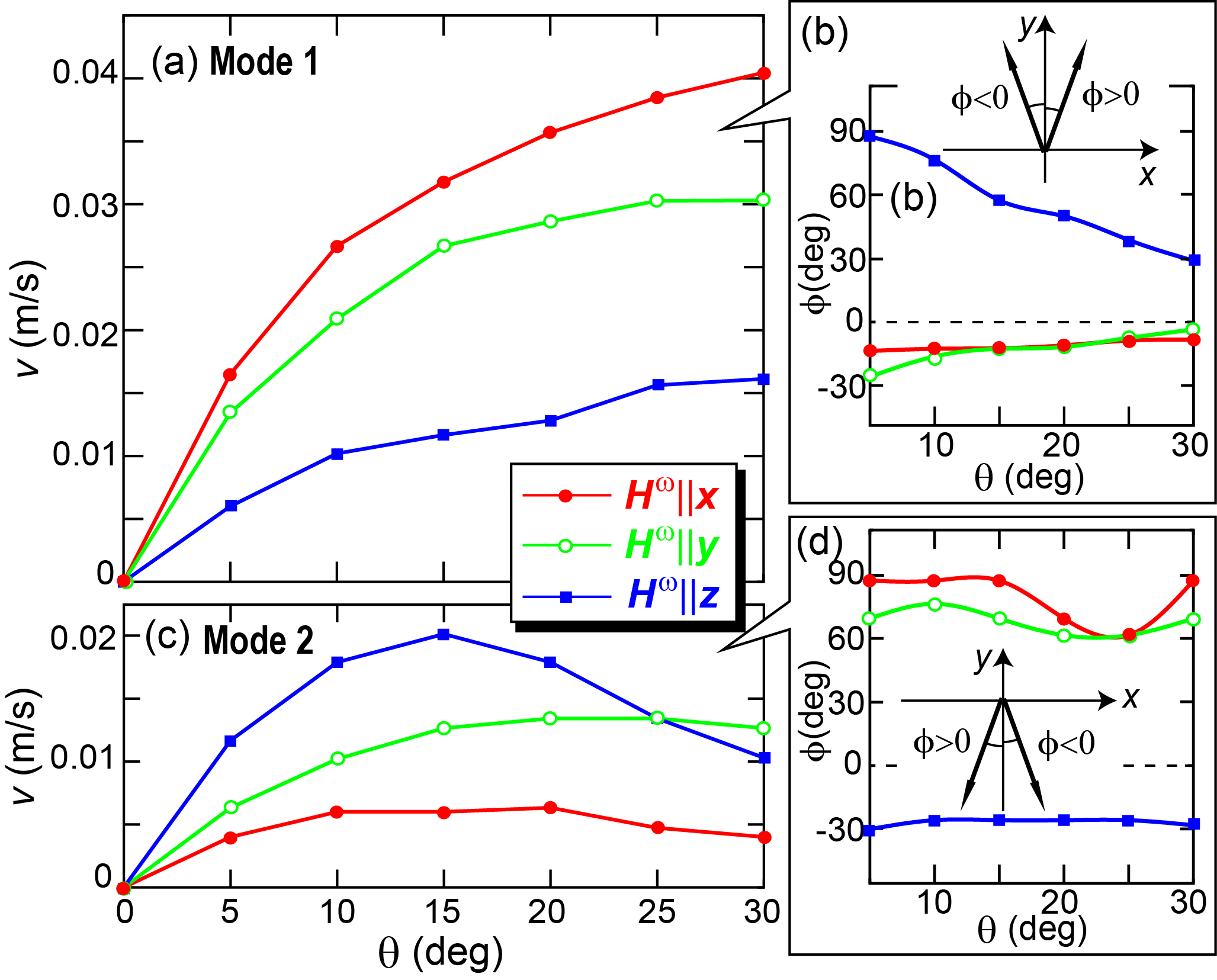}
\caption{(color online). (a) [(c)] Calculated $\theta$-dependence of the velocity $v$=$\sqrt{v_x^2+v_y^2}$ of the skyrmion crystal driven by a microwave field $\bm H^\omega$ through activation of Mode 1 with the dominant counterclockwise-rotation component [Mode 2 with breathing oscillations] for different microwave polarizations. Here $\theta$ is the inclination angle of the external magnetic field $\bm H_{\rm ex}$=$(H_z\tan\theta, 0, H_z)$ with $H_z$=0.036, and the microwave field is given by $H^\omega_\alpha \sin\omega t$ ($\alpha=x,y$) with $H^\omega_\alpha$=0.0006. All velocity data are measured at the resonant frequencies of the modes. (b) [(d)] Calculated $\theta$-dependence of the propagation direction of the skyrmion crystal driven by Mode 1 [Mode 2] for different microwave polarizations.}
\label{Fig07}
\end{figure}
We then investigate $\theta$-dependence of the velocity of the skyrmion crystal when driven by Mode 1 and Mode 2, where $\theta$ is the inclination angle of the external magnetic field $\bm H_{\rm ex}$=$(H_z\tan\theta, 0, H_z)$ with $H_z$=0.036. In Fig.~\ref{Fig07}(a), the calculated absolute values of velocity $v$=$\sqrt{v_x^2+v_y^2}$ are plotted for Mode 1. The velocity increases noticeably as $\theta$ increases and seems to become saturated. Figure~\ref{Fig07}(b) shows the direction of propagation for different microwave polarizations. We find that the skyrmion crystal under the in-plane microwave fields $\bm H^\omega$$\parallel$$\bm x$,$\bm y$ moves approximately in the positive $y$ direction, irrespective of the value of $\theta$. In contrast, the skyrmion crystal under the out-of-plane microwave field $\bm H^\omega$$\parallel$$\bm z$ moves approximately in the positive $x$ direction when $\theta$ is small, whereas the propagation direction becomes slanted towards the positive $y$ direction as $\theta$ increases. 

Figure~\ref{Fig07}(c) shows calculated speeds of $v$=$\sqrt{v_x^2+v_y^2}$ for Mode 2 activated under different microwave polarizations. For $\bm H^\omega$$\parallel$$\bm z$, the velocity initially increases as $\theta$ increases to reach a maximum at $\theta$$\sim$15$^\circ$ and, subsequently, decreases gradually with increasing $\theta$. In contrast, this type of peak-maximum behavior is not clear in the $\bm H^\omega$$\parallel$$\bm x$,$\bm y$ case, but the velocity shows saturation behavior or slight decrease after the initial increase in the small $\theta$ region. Figure~\ref{Fig07}(d) indicates that the direction of propagation under the in-plane microwave fields $\bm H^\omega$$\parallel$$\bm x$,$\bm y$ is approximately in the negative $x$ direction. In contrast, the direction of propagation under the out-of-plane microwave field $\bm H^\omega$$\parallel$$\bm z$ is approximately in the negative $y$ direction and is slanted slightly towards the positive $x$ direction. In both cases, the changes in the propagation direction upon variation of $\theta$ are small.

\begin{figure}
\includegraphics[width=1.0\columnwidth]{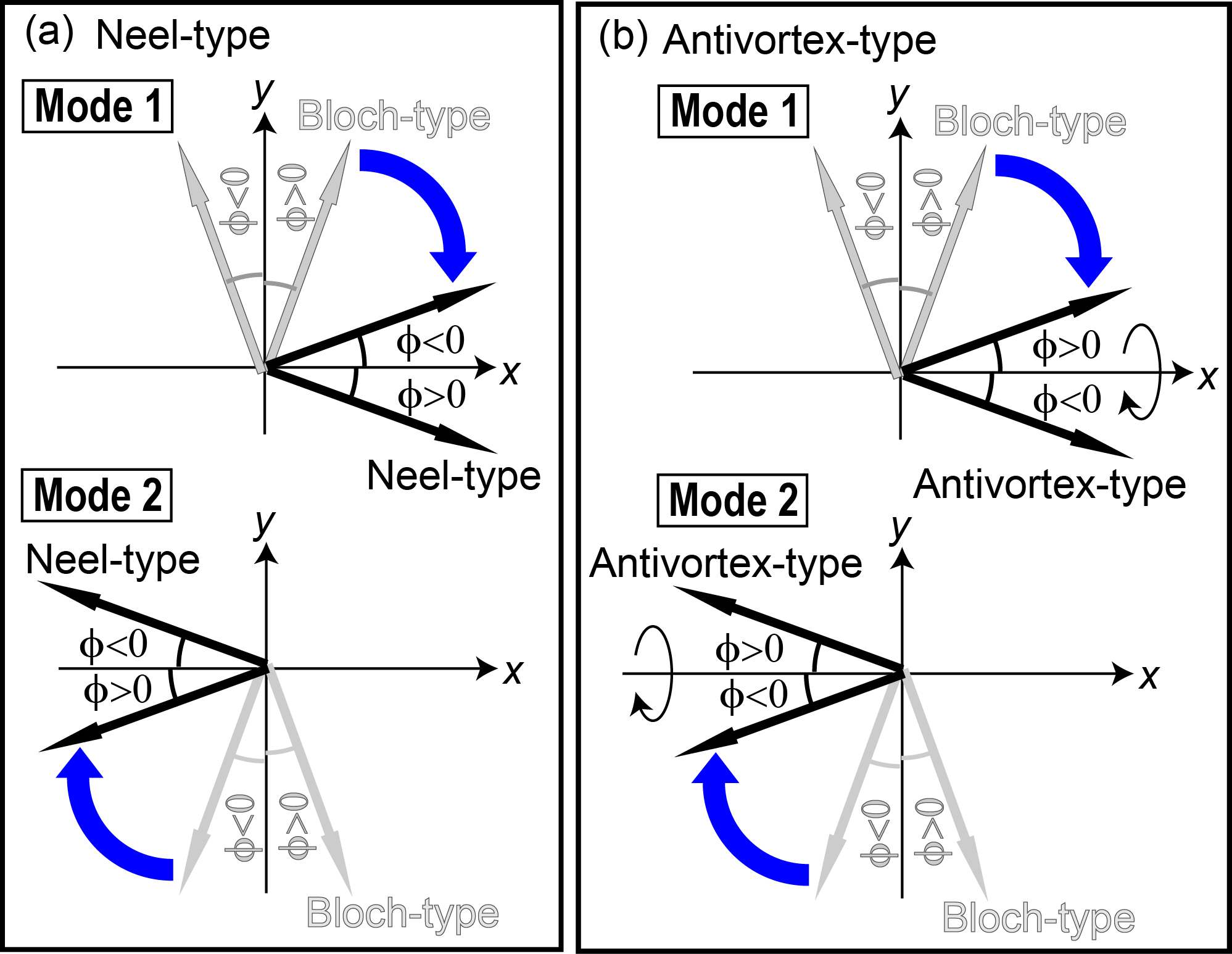}
\caption{(color online). (a) [(b)] Definitions of the angle $\phi$ that describes the direction of propagation of the Neel-type [antivortex-type] skyrmion crystal when driven by Modes 1 and 2. The $\theta$-dependence of the direction of propagation for the Bloch-type skyrmion crystal in Fig.~\ref{Fig07}(b) also holds for the Neel-type and antivortex-type skyrmion crystals if we replace the definitions of $\phi$ with them. Note that they are related to each other through 90$^\circ$ rotation around the $z$ axis and mirror operation with respect to the $zx$ plane.}
\label{Fig08}
\end{figure}
Note that each of the velocity data are measured at the resonant frequency of the mode, which varies depending on the inclination angle $\theta$. The $\theta$-dependence of the resonant frequency $\omega_{\rm R}$ for each mode is summarized in Fig.~\ref{Fig02}(e). It should also be noted that the data shown in Fig.~\ref{Fig07} are calculated for the Bloch-type skyrmion crystal, but we have examined the other two types of skyrmion crystals as well and have found that the absolute speed data plotted in Figs.~\ref{Fig07}(a) and (c) do not alter. In contrast, the direction of propagation changes depending on the type of skyrmion, but they are related to each other. The plots in Fig.~\ref{Fig07}(b) and (d) also hold for the Neel-type and antivortex-type skyrmion crystals if we replace the definitions of $\phi$ in the insets. These definitions should be replaced with those in Fig.~\ref{Fig08}(a) for the Neel-type skyrmion crystal and those in Fig.~\ref{Fig08}(b) for the antivortex-type skyrmion crystal.

\begin{figure}
\includegraphics[width=1.0\columnwidth]{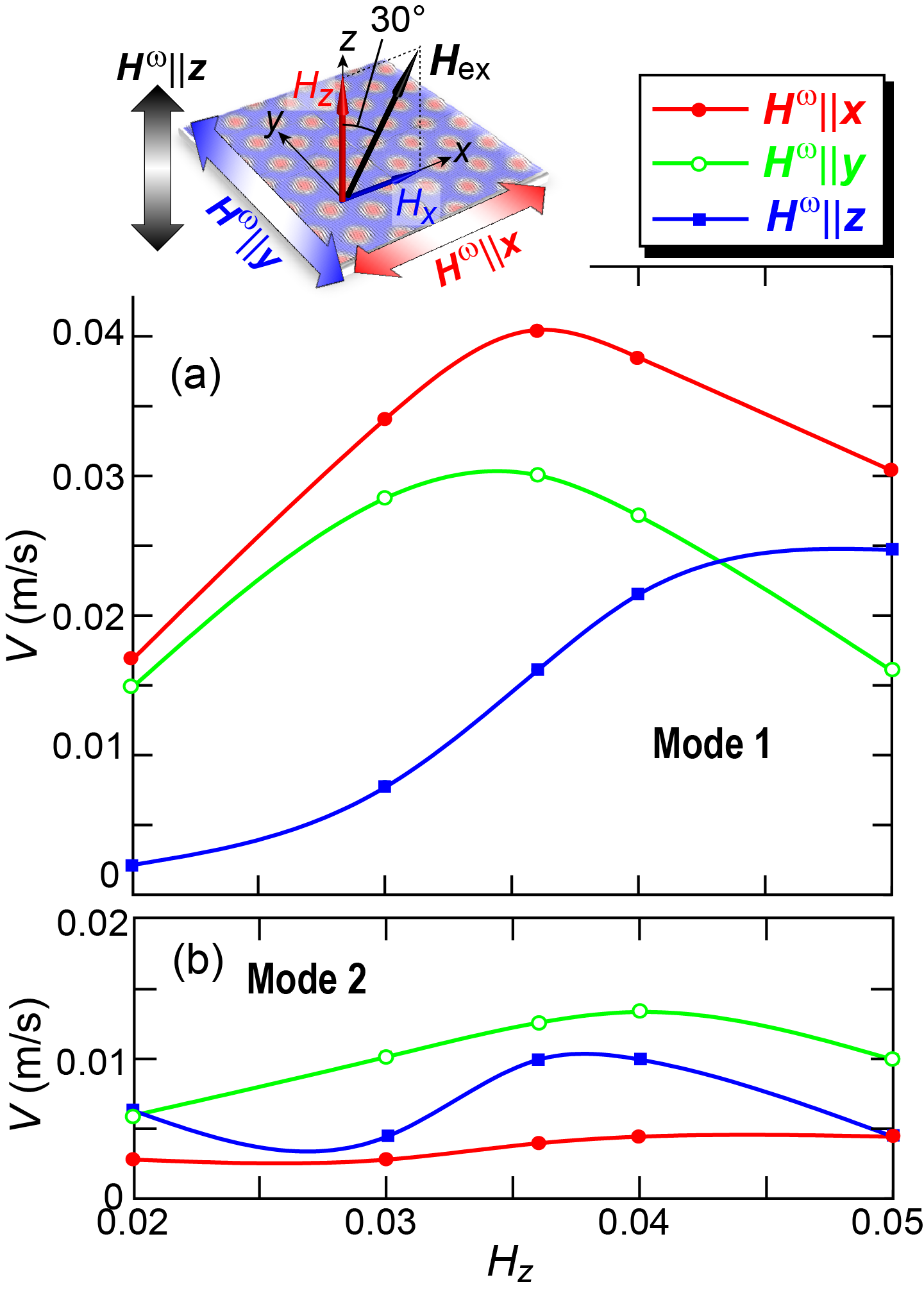}
\caption{(color online). (a) [(b)] Calculated $H_z$-dependence of the velocity $v$=$\sqrt{v_x^2+v_y^2}$ of skyrmion crystal when driven by a microwave field $\bm H^\omega$ through activation of Mode 1 with the dominant counterclockwise-rotation component [Mode 2 with breathing oscillations] for different microwave polarizations. Here $H_z$ is the perpendicular component of the external magnetic field $\bm H_{\rm ex}$=$(H_z\tan\theta, 0, H_z)$ with $\theta$=30$^\circ$, while the microwave field is given by $H^\omega_\alpha \sin\omega t$ ($\alpha=x,y$) with $H^\omega_\alpha$=0.0006.}
\label{Fig09}
\end{figure}
Finally, we examine the $H_z$-dependence of the velocity. Figure~\ref{Fig09}(a) shows the calculated absolute velocity $v$=$\sqrt{v_x^2+v_y^2}$ of skyrmion crystal when driven by a microwave field $\bm H^\omega$ through activation of Mode 1. Here $H_z$ is a perpendicular component of the inclined magnetic field $\bm H_{\rm ex}$=$(H_z\tan\theta, 0, H_z)$ with $\theta$=30$^\circ$. The microwave amplitude is again fixed at $H^\omega_\alpha$=0.0006. The plots for $\bm H^\omega$$\parallel$$\bm x$,$\bm y$ show a maximum at $H_z\sim$0.35, which deep inside the skyrmion crystal phase, whereas the plot for $\bm H^\omega$$\parallel$$\bm z$ monotonically increases with increasing $H_z$. In contrast, the plots for Mode 2 in Fig.~\ref{Fig09}(b) show more complex behavior that is dependent on the microwave polarization. Clarification of a microscopic mechanism of these characteristic behaviors is an issue of importance, which should be clarified in a future.

Note that the present study is based on a pure two-dimensional model. The absolute velocities may slightly vary depending on thickness of real thin-plate specimens because the relative weight of influences from the surface magnetic structures and pinning effects due to impurities, grains and defects must differ depending on the sample thickness. A quantitative investigation on the skyrmion velocities in real three-dimensional specimens is left for future studies.

\section{Summary}
In summary, we have theoretically studied the microwave-active resonance modes of skyrmion crystal on a thin-plate specimen under a perpendicular or inclined external magnetic field $\bm H_{\rm ex}$. We have found that while only two rotation modes or a single breathing mode are active under the perpendicular field, novel microwave-active modes emerge when the $\bm H_{\rm ex}$ field is inclined. The modes that exists originally under the perpendicular field are enhanced or suppressed depending on the polarization of the microwave field $\bm H^\omega$. Recent studies revealed that the collective modes of skyrmions host rich physical phenomena and may provide potentially useful device functions, which are attracting a great deal of research interest. As an example of these phenomena, we have investigated the microwave-driven translational motion of a skyrmion crystal under an inclined $\bm H_{\rm ex}$ field~\cite{WangW15}. Our numerical simulations have demonstrated that the propagation velocity of skyrmion crystal is enhanced at resonant frequencies of the modes, while the velocity and direction of the skyrmion propagation are sensitively dependent on the activated modes. Importantly, the in-plane microwave field that activates the dominant counterclockwise rotation drives the skyrmion crystal much rapidly than the out-of-plane microwave field that activates the breathing mode studied in Ref.~\cite{WangW15}. The knowledge obtained in this study will help to open novel research into the spintronics and magnonics functions of skyrmions based on microwave irradiation.

This work was supported by JSPS KAKENHI (Grant No. 17H02924), Waseda University Grant for Special Research Projects (Project Nos. 2017S-101, 2018K-257), and JST PRESTO (Grant No. JPMJPR132A).


\begin{thebibliography}{999}
\bibitem{Bogdanov89}A. N. Bogdanov and D.A. Yablonskii, Sov. Phys. JETP {\bf 68}, 101 (1989).

\bibitem{Bogdanov94}A. Bogdanov and A. Hubert, J. Mag. Mag. Mat. {\bf 138}, 255 (1994).

\bibitem{Rossler06}U.K. R\"o{\ss}ler, A. N. Bogdanov, and C. Pfleiderer, Nature {\bf 442}, 797 (2006).

\bibitem{Nagaosa13}N. Nagaosa and Y. Tokura, Nat. Nanotech. {\bf 8}, 899 (2013).
\bibitem{Fert13}A. Fert, V. Cros, and J. Sampaio, Nat. Nanotech. {\bf 8}, 152 (2013).

\bibitem{Mochizuki15a}M. Mochizuki and S. Seki, J. Phys.: Cond. Matt. {\bf 27}, 503001 (2015).

\bibitem{Seki15}S. Seki and M. Mochizuki, ``Skyrmions in Magnetic Materials" (Springer Briefs in Physics).

\bibitem{Muhlbauer09}S. M\"uhlbauer, B. Binz, F. Jonietz, C. Pfleiderer, A. Rosch, A. Neubauer, R. Georgii, and P. B\"oni, Science {\bf 323}, 915 (2009).

\bibitem{YuXZ10}X. Z. Yu, Y. Onose, N. Kanazawa, J. H. Park, J. H. Han, Y. Matsui, N. Nagaosa, and Y. Tokura, Nature {\bf 465}, 901 (2010).

\bibitem{Seki12}S. Seki, X. Z. Yu, S. Ishiwata, and Y. Tokura, Science {\bf 336}, 198 (2012).

\bibitem{Adams12}T. Adams, A. Chacon, M. Wagner, A. Bauer, G. Brandl, B. Pedersen, H. Berger, P. Lemmens, and C. Pfleiderer, Phys. Rev. Lett. {\bf 108}, 237204 (2012).

\bibitem{Tonomura12}A. Tonomura X. Z. Yu, K. Yanagisawa, T. Matsuda, Y. Onose, N. Kanazawa, H. S. Park, and Y. Tokura, Nano Lett. {\bf 12}, 1673 (2012).

\bibitem{Mochizuki12}M. Mochizuki, Phys. Rev. Lett. {\bf 108}, 017601 (2012).

\bibitem{Petrova11}O. Petrova and O. Tchernyshyov, Phys. Rev. B {\bf 84}, 214433 (2011).

\bibitem{LinSZ14}S.-Z. Lin, C. D. Batista, and A. Saxena, Phys. Rev. B {\bf 89}, 024415 (2014).

\bibitem{Schwarze15}T. Schwarze, J. Waizner, M. Garst, A. Bauer, I. Stasinopoulos, H. Berger, C. Pfleiderer, and D. Grundler, Nat. Mater. {\bf 14}, 478 (2015).

\bibitem{Garst17}M. Garst, J. Waizner, and D. Grundler, J. Phys. D: Appl. Phys. {\bf 50}, 293002 (2017).

\bibitem{Finocchio16}G. Finocchio, F. B\"uttner, R. Tomasello, M. Carpentieri, and M. Kl\"aui, J. Phys. D: Appl. Phys. {\bf 49}, 423001 (2016).
\bibitem{Mochizuki13}M. Mochizuki, and S. Seki, Phys. Rev. B {\bf 87}, 134403 (2013).

\bibitem{Okamura13}Y. Okamura, F. Kagawa, M. Mochizuki, M. Kubota, S. Seki, S. Ishiwata, M. Kawasaki, Y. Onose, Y. Tokura, Nat. Commun. {\bf 4}, 2391 (2013).

\bibitem{Mochizuki15b}M. Mochizuki, Phys. Rev. Lett. {\bf 114}, 197203 (2015).

\bibitem{Okamura15}Y. Okamura, F. Kagawa, S. Seki, M. Kubota, M. Kawasaki, and Y. Tokura, Phys. Rev. Lett. {\bf 114}, 197202 (2015).
\bibitem{Ohe13}J. Ohe and Y. Shimada, Appl. Phys. Lett. {\bf 103}, 242403 (2013).

\bibitem{Shimada15}Y. Shimada and J. I. Ohe, Phys. Rev. B {\bf 91}, 174437 (2015).
\bibitem{Hirobe15}D. Hirobe, Y. Shiomi, Y. Shimada, J. Ohe, and E. Saitoh, J. Appl. Phys. {\bf 117} 053904 (2015).
\bibitem{LiuRH15}R. H. Liu, W. L. Lim, and S. Urazhdin, Phys. Rev. Lett. {\bf 114}, 137201 (2015).

\bibitem{ZhangS15}S. Zhang, J. Wang, Q. Zheng, Q. Zhu, X. Liu, S. Chen, C. Jin, Q. Liu, C. Jia, and D. Xue, New J. Phys. {\bf 17}, 023061 (2015).
\bibitem{Finocchio15}G. Finocchio, M. Ricci, R. Tomasello, A. Giordano, M. Lanuzza,  V. Puliafito,  P. Burrascano, B. Azzerboni, and  M. Carpentieri, Appl. Phys. Lett. {\bf 107}, 262401 (2015).
\bibitem{MaF15}F. Ma, Y. Zhou, H. B. Braun, and W. S. Lew, Nano Lett. {\bf 15}, 4029 (2015).

\bibitem{MoonKW16}K.-W. Moon, B. S. Chun, W. Kim, and C. Hwang, Phys. Rev. Applied {\bf 6}, 064027 (2016).

\bibitem{Mruczkiewicz16}M. Mruczkiewicz, P. Gruszecki, M. Zelent, and M. Krawczyk, Phys. Rev. B {\bf 93}, 174429 (2016).

\bibitem{YuXZ11}X. Z. Yu, N. Kanazawa, Y. Onose, K. Kimoto, W. Z. Zhang, S. Ishiwata, Y. Matsui, and Y. Tokura, Nat. Mater. {\bf 10} 106 (2011).

\bibitem{LinSZ15}S.-Z. Lin and A. Saxena, Phys. Rev. B {\bf 92}, 180401(R) (2015).
\bibitem{WangW15}W. Wang, M. Beg, B. Zhang, W. Kuch, and H. Fangohr, Phys. Rev. B {\bf 92}, 020403(R) (2015).
\bibitem{Kezsmarki15}I. K\'ezsm\'arki, S. Bord\'acs, P. Milde, E. Neuber, L. M. Eng, J. S. White, H. M. R$\o$nnow, C. D. Dewhurst, M. Mochizuki, K. Yanai, H. Nakamura, D. Ehlers, V. Tsurkan, and A. Loidl, Nat. Mater. {\bf 14}, 1116 (2015).

\bibitem{Ehlers16}D. Ehlers, I. Stasinopoulos, V. Tsurkan, H.-A. Krug von Nidda, T. Feh\'er, A. Leonov, I. K\'ezsm\'arki, D. Grundler, and A. Loidl, Phys. Rev. B {\bf 94} 014406 (2016).

\bibitem{Ehlers17}D. Ehlers, I. Stasinopoulos, I. K\'ezsm\'arki, T. Feh\'er, V. Tsurkan, H.-A. Krug von Nidda, D. Grundler, and A. Loidl, J. Phys.: Condens. Matter {\bf 29} 065803 (2017).
\bibitem{Bak80}P. Bak, and M. H. Jensen, J. Phys. C {\bf 13}, L881 (1980).

\bibitem{YiSD09}S. D. Yi, S. Onoda, N. Nagaosa, and J. H. Han, Phys. Rev. B {\bf 80}, 054416 (2009).
%
\bibitem{Kim14}J.-V. Kim, F. Garcia-Sanchez, J. Sampaio, C. Moreau-Luchaire, V. Cros, and A. Fert, Phys. Rev. B {\bf 90}, 064410 (2014).

\bibitem{Beg17}M. Beg, M. Albert, M.-A. Bisotti, D. Cort\'es-Ortu\~no, W. Wang, R. Carey, M. Vousden, O. Hovorka, C. Ciccarelli, C. S. Spencer, C. H. Marrows, and H. Fangohr, Phys. Rev. B {\bf 95}, 014433 (2017).

\bibitem{Kumar11}D. Kumar, O. Dmytriiev, S. Ponraj, and A. Barman, J. Phys. D: Appl. Phys. {\bf 45}, 015001 (2011).
\bibitem{Nayak17}A. K. Nayak, V. Kumar, T. Ma, P. Werner, E. Pippel, R. Sahoo, F. Damay, U. K. Rosler, C. Felser, and S. S. P. Parkin, Nature {\bf 548}, 561 (2017).

\bibitem{Rybakov16}F. N. Rybakov, A. B Borisov, S. Bl\"ugel, and N. S. Kiselev, New J. Phys. {\bf 18}, 045002 (2016).

\bibitem{ZhangSL18}S. L. Zhang, G. van der Laan, W. W. Wang, A. A. Haghighirad, and T. Hesjedal, Phys. Rev. Lett. {\bf 120}, 227202 (2018).
\end{thebibliography}
\end{document}